\documentclass[twocolumn,preprintnumbers,amsmath,amssymb]{revtex4-1}
\usepackage{epsfig}
\usepackage{amsmath}
\begin{document}
\title{Free energy calculations for atomic solids through the Einstein crystal/molecule methodology using GROMACS and LAMMPS}
\author{J.L. Aragones, C. Valeriani and C. Vega}
\affiliation{Departamento de Qu\'{\i}mica F\'{\i}sica,
Facultad de Ciencias Qu\'{\i}micas, Universidad Complutense de Madrid,
28040 Madrid, Spain}
\date{\today}
\begin{abstract}
In this work the free energy of solid phases is computed for the Lennard-Jones potential and for a model of NaCl.
The free energy is evaluated through the Einstein crystal/molecule methodologies using the Molecular
Dynamics programs: GROMACS and LAMMPS. The obtained results are compared with the results obtained from Monte 
Carlo. Good agreement between the different programs and methodologies was found. 
The procedure to perform the free energy calculations for the solid phase in 
the Molecular Dynamic programs is described. Since these programs allow to study any continuous intermolecular potential
(when given in a tabulated form) this work shows that for isotropic potentials (describing for instance
atomic solids or colloidal particles) free energy calculations can be performed
on a routinely basis using GROMACS and/or LAMMPS.  
\end{abstract}

\maketitle

\twocolumngrid

In 1984 Frenkel and Ladd proposed the ``Einstein crystal method'', 
a novel scheme to compute the free energy of a solid~\cite{JCP_1984_81_03188}: 
the method is based on a thermodynamic integration of the Helmholtz free energy 
in the canonical ensemble along a reversible path between the system of interest and an ideal Einstein crystal with the same structure as the real 
solid, whose Helmholtz free energy can be analytically computed.
In the ideal Einstein crystal, particles are attached to their lattice 
positions via harmonic springs (of constant $\Lambda_E$).
More recently, some of us have proposed the 
``Einstein molecule method''~\cite{vega_noya,noya_conde_vega}, a variant of the 
Einstein crystal to compute the free energy of molecular solids.
The main difference between both methods is the choice of the reference system. 
In the Einstein crystal the reference system is an ideal Einstein 
crystal with the constraint of the center-of-mass of the system
kept fixed (to avoid a quasi-divergence of the integral 
of the free energy change from the reference system to the real solid). 
In the Einstein molecule the reference system is an 
ideal Einstein crystal with the constraint of 
the position of one particle kept fixed. 

The Helmholtz free energy $A_{sol}(T,V)$ computed with the Einstein crystal/molecule
calculations can be written as $A_{sol}(T,V) =  A_0(T,V) +  \Delta A_1(T,V)  +  \Delta A_2(T,V)$, 
where A$_0$ is the free energy of the reference system (including corrections for the fixed point),
whose analytical expression is slightly
different in the Einstein crystal and Einstein molecule, 
$\Delta$A$_1$ is the free energy difference between the ideal Einstein crystal
and the Einstein crystal in which particles interact through the Hamiltonian of the real solid
(``interacting'' Einstein crystal) and $\Delta$A$_2$  is  the free energy difference between 
the interacting Einstein crystal and the real solid~\cite{vega_review}.
The expression for both $\Delta$A$_1$ and $\Delta$A$_2$ is the same in both methods 
although its value may be different since the fixed point (center-of-mass or a reference particle) is different
(a detailed description of these terms is provided as Supplementary Material\cite{material_suplementario}).
In any case since the free energy of a solid is uniquely defined, its value  
should not depend on the methodology used to compute it~\cite{vega_noya,vega_review}.

Besides the Einstein crystal methods other methods have been developed in the last decade 
to estimate 
the free energy of solids (or to determine the fluid-solid equilibria), such as 
the Phase Switch method~\cite{bruce97} and an integration path 
to calculate the Gibbs free energy difference 
between any arbitrary solid and liquid proposed by 
Grochola~\cite{JCP_2004_120_02122}. 
In spite of this progress the number of groups performing free energy calculations for 
solids is still small\cite{JCP_2003_118_00728,ACP_2000_115_0113_nolotengo,hynninen}.
Solid free energy calculations can be easily adapted into Monte Carlo scheme,
but it requires one to write {\it a la carte} MC codes. Therefore it seems of interest to consider
if these calculations can be implemented in widely used open source Molecular Dynamics simulation programs 
such as GROMACS~\cite{spoel05} or LAMMPS~\cite{lammps}. 
As far as we are aware, free energy calculations of solids are not yet 
explicitly implemented in such packages and only recently 
free energy calculations for alloys of copper and zirconium\cite{harrowell} using LAMMPS through the Einstein crystal method have been reported.  
Since  GROMACS and/or LAMMPS incorporate the possibility of using harmonic
potentials restraining the position of atoms (or even freezing the position of atoms) and 
incorporate thermostats that can correctly treat harmonic oscillators (as for instance 
the new thermostat of Bussi et al.\cite{JCP_126_014101_2007}), it seems possible to perform Einstein crystal/molecule
calculations for atomic solids using these programs.
In this note we shall use  GROMACS and LAMMPS to compute the
free energy of solids of particles interacting via isotropic potentials 
using either the Einstein crystal or the Einstein molecule method.
In particular we will compute the free energy of Lennard-Jones (LJ)
 and of Sodium Chloride modeled via the Joung-Cheatham(JC)-NaCl potential~\cite{JPCB_112_9020_2008} (using the SPC/E set) 
which describes ion-ion interactions through a LJ potential plus
a Coulomb term. 
To confirm the results obtained with  
GROMACS (G) and LAMMPS (L), calculations will be also performed using a Monte Carlo code (MC).
\begin{table}[h!]
\centering
\resizebox{8.5cm}{!}{
\begin{tabular}{llcccccc}
\hline
  &  & N  &  r$_c$         & $A_0$ & $\Delta A_1$ & $\Delta A_2$ & $A_{sol}$ \\
\hline
\hline
{\bf EC} & & & & & & \\
\hline
MC & LJ/STS  & 256 & $2.7\sigma$ & 13.61 & -3.14(1) & -7.36(3) & 3.11(4) \\
G  & LJ/STS  & 256 & $2.7\sigma$ & 13.61 & -3.14(1) & -7.35(3) & 3.10(4) \\
L  & LJ/STS  & 256 & $2.7\sigma$ & 13.61 & -3.14(1) & -7.36(3) & 3.11(4) \\
\hline
MC & LJ/ST   & 1372 & $5\sigma$ & 13.68 & -3.69(1) & -7.40(3) & 2.59(4) \\
G  & LJ/ST   & 1372 & $5\sigma$ & 13.68 & -3.69(1) & -7.38(3) & 2.61(4) \\
\hline
MC & JC-NaCl  & 1000 & 14~\AA & 10.70 & -159.94(1) & -6.34(4) & -155.58(5) \\
G  & JC-NaCl  & 1000 & 14~\AA & 10.70 & -159.95(1) & -6.30(4) & -155.55(5) \\
\hline
\hline
{\bf EM} & & & & & & \\
\hline
MC & LJ/STS  & 256 & $2.7\sigma$ & 13.64 & -3.14(1) & -7.39(3) & 3.11(4) \\
G  & LJ/STS  & 256 & $2.7\sigma$ & 13.64 & -3.14(1) & -7.38(3) & 3.12(4) \\
L  & LJ/STS  & 256 & $2.7\sigma$ & 13.64 & -3.15(1) & -7.40(3) & 3.09(4) \\
\hline
MC & LJ/ST   & 1372 & $5\sigma$ & 13.69 & -3.69(1) & -7.40(3) & 2.60(4) \\
G  & LJ/ST   & 1372 & $5\sigma$ & 13.69 & -3.69(1) & -7.39(3) & 2.61(4) \\
\hline
MC & JC-NaCl & 1000 & 14~\AA & 10.71 & -159.95(1) & -6.35(4) & -155.59(5) \\
G  & JC-NaCl & 1000 & 14~\AA & 10.71 & -159.95(1) & -6.34(4) & -155.57(5) \\
\hline
\end{tabular}
}
\caption{
Free-energies of solids as obtained from Einstein Crystal (EC) and Einstein Molecule (EM) methodologies
using MC or MD (G;L). N is the number of particles, $r_c$ is the cut-off distance of the LJ contribution.
Results for the LJ/STS and LJ/ST systems were obtained at  $T^*=k_BT/\epsilon=2$ and $\rho^*=\rho \sigma^3=1.28$ for the fcc 
structure. Results for the NaCl solid were obtained for the JC-NaCl model at $T=298~K$ and a volume of $V=24.13$~nm$^3$.
Free energies are in $Nk_BT$ units. 
\label{free_energies}}
\end{table}

We have carried out free energy calculations for a 256 LJ-Argon
spherically truncated and shifted (STS)
at $r_c=2.7\sigma$ and for a 1372 LJ spherically truncated (ST) at $r_c=5\sigma$.
Results of the free energy calculations are presented in Table~\ref{free_energies}.
For the LJ/STS system  A$_0$, $\Delta$ A$_1$, and $\Delta$ A$_2$ 
obtained from MC and from MD (GROMACS/LAMMPS) are in very good agreement, with 
 a free energy difference lower than 0.03~$Nk_BT$ (typical uncertainties in calculations of
free energy of solids being of 0.05~$Nk_BT$).
The choice of a LJ spherically truncated and shifted (STS) avoids the subtle issues arising when  comparing
results obtained by MC and MD for a spherically truncated (ST) 
potential\cite{material_suplementario,demiguel06a,frenkel_smit}.
In the LJ/ST truncated at $r_c=5\sigma$, the free energy results obtained with MC and
MD agree quite well with each other and with previous calculations
for the same system size and thermodynamic state~\cite{vega_noya,barroso,JCP_136_144508_2012}.
It is clear that the magnitude of problems arising with the discontinuity of the potential at the cut-off are also quite
small when the cut-off distance is large enough ($\approx 5\sigma$).

Notice that although for a certain system size and Hamiltonian the computed free energy of a solid is unique, 
in general the free energy of solids changes with the system size.
As a suggestion, in order to have a reasonable estimate of the free energy of a solid,
we recommend to use a relatively large system size
(with more than 1000 molecules) and to average properties over around $10^4-10^5$ independent configurations
for evaluating $\Delta A_1$ and runs of about 10~ns for $\Delta A_2$.
Moreover,  the use of a large cut-off 
(above $4.5\sigma$) is recommended.
In all LJ systems, both Einstein molecule and Einstein crystal
give the same value of the free energy $A_{sol}$, although the single 
contributions are slightly different in both methodologies (as they should be). 
In the Supplementary Material, we provide further details on the implementation\cite{material_suplementario},  
needed for GROMACS and LAMMPS 
to easily compute A$_0$, $\Delta$ A$_1$, and $\Delta$ A$_2$.

To show that the methodology also works for more complex systems
where particles interact via an isotropic potential, we calculate
the free energy of NaCl using the Joung-Cheatham potential~\cite{JPCB_112_9020_2008}. 
Results are presented in Table~\ref{free_energies}. 
We simulate a 1000 ions NaCl solid (at 298~K and the equilibrium density of the model at 1~bar). 
The free energies computed from MC and GROMACS agree reasonably well
(the difference been of 0.03~$Nk_BT$) and 
are in good agreement with previous calculations~\cite{aragones_nacl}.

To conclude, we have demonstrated that it is possible to compute the free energy of
atomic solids using GROMACS and LAMMPS for systems interacting with spherical potentials
such as Lennard-Jones, Yukawa, Morse and any continuous potential since both packages incorporate the possibility of using a tabulated numerical intermolecular potentials. We do hope that this work stimulate more 
groups to perform free energy calculations for solids. 
For future work it would be useful to analyze if for molecular fluids the Einstein crystal/molecule calculations 
could also be performed with GROMACS and/or LAMPPS. 

This work was funded by the grants FIS2010-16159 (DGI) and MODELICO-P2009/ESP/1691 (CAM).
J.L. Aragones acknowledges MEC for a pre-doctoral grant. 
C. Valeriani acknowledges financial support from a Juan de la Cierva 
Fellowship and from the Marie Curie Integration Grant PCIG-GA-2011-303941 ANISOKINEQ. 

\bibliographystyle{./apsrev}

\end{document}


\title{Supplementary information: Free energy calculations for atomic solids through the Einstein crystal/molecule methodology using GROMACS and LAMMPS}
\date{\today}

\maketitle

\twocolumngrid


The Helmholtz free energy of a solid can be computed using either the Einstein Crystal~\cite{JCP_1984_81_03188} or
 the Einstein molecule method~\cite{vega_noya}. 
The two mainly differ for the choice of the reference system: 
in the Einstein crystal the reference system is an ideal Einstein 
crystal where the center of mass is kept fixed, whereas 
in the Einstein molecule the reference system is an 
ideal Einstein crystal where one particle is kept fixed.

The Helmholtz free energy ($A_{sol}$) of a solid can be written as the sum of three terms:
\begin{equation}
\label{tot}
A_{sol}(T,V) =  A_0(T,V) +  \Delta A_1(T,V)  +  \Delta A_2(T,V), 
\end{equation}
where A$_0$ is the free energy of the ideal Einstein crystal reference system
(including corrections for the fixed point), 
$\Delta$A$_1$ is the free energy difference between the ideal Einstein crystal
and the Einstein crystal in which particles interact through the Hamiltonian of the solid 
of interest (interacting Einstein crystal) and $\Delta$A$_2$ is the the free energy difference between 
the interacting Einstein crystal and the solid of interest~\cite{vega_review}.

\subsection{Analytical calculation of A$_0$}
The analytical calculation of A$_0$ is different for the Einstein crystal and the Einstein molecule 
as depends on the chosen reference system. 
In the Einstein crystal method, A$_0$ contains the analytical free energy of an ideal
Einstein crystal with fixed  center-of-mass and the free energy difference between
the solid with and without the fixed center-of-mass~\cite{vega_review}:
\begin{equation}
\label{ECa0}
\frac{A_0^{EC}}{Nk_{B}T}=-\frac{1}{N}\ln\left[ \frac{1}{\Lambda^{3N}} \left(\frac{\pi}{\beta \Lambda_E}\right)^{3(N-1)/2}N^{3/2} \frac{V}{N} \right]
\end{equation}
where $N$ is the number of particles, $\Lambda$ the thermal De Broglie wave length, 
$\beta=1/k_BT$ (with $k_B$ the Boltzmann constant), $V$  the system's volume  and $\Lambda_E$ the 
harmonic spring constant. It can be rewriten as:
\small
\begin{equation}
\label{ECa0b}
\frac{A_0^{EC}}{Nk_{B}T}=\frac{3}{2}\left(1-\frac{1}{N}\right)\ln\left(\frac{\beta \Lambda_E \Lambda^2}{\pi}\right)+\frac{1}{N}\ln\left(\frac{N \Lambda^3}{V}\right)-\frac{3}{2 N}\ln\left(N\right)
\end{equation}
\normalsize
Whereas for the Einstein molecule the expression for $A_0$ is:
\begin{equation}
\label{EMa0}
\frac{A_0^{EM}}{Nk_{B}T}=\frac{3}{2}\left(1-\frac{1}{N}\right)\ln\left(\frac{\beta \Lambda_E \Lambda^2}{\pi}\right)+\frac{1}{N}\ln\left(\frac{N \Lambda^3}{V}\right)
\end{equation}
The absolute value of $A_0$ ( and therefore of $A_{sol}$) depends on the value assigned
to the thermal de Broglie wave length. However, phase coexistence properties do not depend on this value provided that the
same value of $\Lambda$ is used in all phases. This just reflects the fact that in classical statistical thermodynamics
the values of the masses do not affect coexistence properties. For this reason it is a common practice to set the
value of $\Lambda$ to an arbitrary convenient value. 
In this work we set the value of the De Broglie thermal wave length ($\Lambda$) 
to $\sigma$ for the LJ systems and to 1~\AA~for NaCl. 
In any case it is always possible to use the correct value of
$\Lambda= h/\sqrt{2 \pi m k_{B} T}  $ in the calculations.

While  A$_0$ can be computed  analytically, numerical simulations are needed to calculate 
 $\Delta A_1$  and  $\Delta A_2$. 
 In this manuscript we show 
that  this calculation can be implemented either with GROMACS or with LAMMPS.

\subsection{Calculation of  $\Delta$ A$_1$  with GROMACS/LAMMPS}

To compute $\Delta$ A$_1$, we suggest the following steps both in GROMACS and in LAMMPS:
\begin{enumerate}

\item We prepare an ideal Einstein crystal with the crystalline structure of the solid of interest,
that means ideal gas particles attached to their lattice positions by harmonic springs.

\item We equilibrate it with GROMACS/LAMMPS fixing either the center-of-mass (Einstein crystal) 
or the position of one particle (Einstein molecule) while letting all particles (Einstein crystal) 
or all other particles (Einstein molecule) vibrating through harmonic springs around their lattice positions.
Both packages allow to "freeze" either the center-of-mass of the system or one particle's position.

\item The simulation is carried out in the NVT ensemble at the temperature
and density of interest, and we recommend to store around $10^4-10^5$ configurations
of the trajectory to properly compute the ensemble average.

\item $\Delta$ A$_1$ is computed as:
\small
\begin{equation}
\label{deltaa1}
\beta\Delta A_{1}  = \beta U_{lattice} - \ln \left<\exp\left[-\beta(U_{sol}-U_{lattice})\right]\right>_{Ein-id}
\end{equation}
\normalsize
where $U_{lattice}$ is the potential energy of the perfect lattice, which can be estimated running GROMACS/LAMMPS
for a perfect lattice using just one MD step and zero as the time step, and $U_{sol}$ is the potential
energy of the instantaneous configuration; evaluated using the intermolecular potential of interest.

\end{enumerate}

Splitting the calculation of  $\Delta A_1$ and  $\Delta A_2$ allows 
to choose the proper value for the harmonic spring constant $\Lambda_E$.
As an empirical rule, we suggest that a good choice of $\Lambda_E$ is the one that leads 
to a value of $\Delta A_1$  approximately 0.02~$Nk_BT$ higher than the lattice energy  $U_{lattice}$.
This procedure is important because the latest release of GROMACS does not incorporate yet the 
Hamiltonian integration, available in the latest release of 
 LAMMPS but only for simple potentials.

\subsection{Calculation of  $\Delta$ A$_2$  with GROMACS/LAMMPS}

We then evaluate  $\Delta$ A$_2$ in a NVT ensemble as 
\begin{equation}
\label{deltaa2}
\Delta A_2 =-\int^{\Lambda_E}_{0}\left<\sum_i^N\left(r_i-r_{i,0}\right)^2\right>_{N,V,T,\Lambda_E'}d\Lambda_E'
\end{equation}
where the integrand is simply the mean square displacement of each particle 
from its lattice position. This term can be easily obtained with GROMACS/LAMMPS 
commands, such as  {\it position-restraint/ fix spring/self }, respectively, 
that apply a spring force on each particle to tether them to their initial position.
From the calculation of the total harmonic energy it is possible to compute the
mean square displacement for each value of the spring constant ($\Lambda_E'$). In fact since 
 $U_{pos-rest}= \Lambda_E' \left<\sum_i^N\left(r_i-r_{i,0}\right)^2\right>_{N,V,T,\Lambda_E'} =  \frac{k'}{2}\left<\sum_i^N\left(r_i-r_{i,0}\right)^2\right>_{N,V,T,k'}$, the mean square displacement is simply obtained by dividing $U_{pos-rest}$ by
either $\Lambda_E'$ or $k'/2$ (notice that $\Lambda_E'$=$k'/2$). 
 The maximum value of $\Lambda_E'$ used in the calculations is 
denoted as $\Lambda_E$ whereas the maximum value of $k'$ used in the calculations is denoted as $k$.
Notice that depending on the program and algorithm one should use  either $\Lambda_E'$ or $k'$ in the input files.
We then compute the integral in Eq.\ref{deltaa2} using the Gaussian quadrature method with 15 values of $\Lambda_E'$ for
the LJ system and 12 for the NaCl system.
When implementing the Einstein molecule with LAMMPS we did not use the fix self spring command
but rather prepared the initial configuration with "ghosts" atoms at the lattice positions,
tethered to the real atoms via harmonic springs with constant  $\Lambda_E'$.

\subsection{Details on the implementation of the Einstein crystal and Einstein molecule with GROMACS/LAMMPS}

When implementing the Einstein crystal method, we need to simulate the system with 
fixed center-of-mass and the latest releases of GROMACS and LAMMPS  can easily keep fixed
the system's center-of-mass. Whereas when implementing the Einstein molecule method,
we simulate the system keeping fixed the position of one particle, that can be freezed both in GROMACS and in LAMMPS: 
this let the center-of-mass freely move, as shown in  Fig.~\ref{com} for MC and GROMACS.
\begin{figure}[h!]
\centering
\includegraphics[clip,scale=0.3,angle=-0]{com.eps}
\caption{Center-of-mass displacement of the LJ/STS system for the spring constant $\Lambda_E'=$27.08~$k_BT$/\AA$^2$
obtained with MC (left-hand side) and GROMACS (right-hand side) obtained in Einstein molecule
calculations. \label{com}}
\end{figure}

We have simulated Lennard-Jones and JC-NaCl systems in a NVT ensemble using the v-rescaling
thermostat\cite{JCP_126_014101_2007} 
and tested that the results obtained do not depend on the chosen relaxation time of the thermostat $\tau$
(up to  $\tau$=2~ps). In order to show that our results are independent on the chosen thermostat, 
we have also simulated the LJ/STS using  the Langevin thermostat~\cite{lange}  
and found no effect on the final calculation
of the free energy of the solid. 
However, we have not equilibrated the system with a Nose-Hoover thermostat~\cite{nose1,nose2}  since 
it presents a pathological behavior with harmonic potentials\cite{nose2}.

It is important to choose the time step carefully when computing  $\Delta$ A$_2$ or $\Delta$ A$_1$.
The  period of an oscillation of an ideal harmonic spring is a function of the spring constant, 
\begin{equation}
\frac{1}{\nu}={2\pi} \sqrt{\frac{m}{2 \Lambda_E'}}
\end{equation}
We use a Molecular Dynamics time step of 2.5 - 5.0~fs,
since that allowed a correct sampling of the vibrations for the strengths of the springs used in this work.


\section{Simulation details}
Free energy calculations for the solid were performed in the NVT ensemble.
We carry out free energy calculations for a spherically truncated 
and shifted (STS) and for spherically truncated (ST) LJ systems
at $T^*=2$ and $\rho^*=1.28$.
Free energy calculations were performed for the fcc solid structure.
For the STS system we used 256 atoms along with the cutoff
distance $r_c= 2.7\sigma$.
For the ST system we used 1372 atoms along with the cutoff distance $r_c=5\sigma$. 
Simulations results for the LJ potentials (both STS and ST) were implemented using a
LJ-Argon system (with
$\sigma=3.405$\AA, $\epsilon/k=120$K and $m=39.9$~g/mol). 
The maximum value of the spring constant was $\Lambda_E=2500 k_BT$ / \AA$^2$ for the LJ systems.
For NaCl the free energy of the solid was obtained from NVT runs for a system containing 1000 ions and
using the Joung-Cheatham-NaCl model (optimized for SPC/E water). 
Calculations were performed at 298~K and $V=24.13$~nm$^3$ ( which correspond
to the average value of the volume of the system for the considered system size and model 
obtained from a previous NpT run at 298~K and 1 bar).
The maximum value of the spring constant for NaCl was  
$\Lambda_E=4000 k_BT$ / \AA$^2$ .
For the NaCl system, Ewald sums were used ({\em Particle Mesh Ewald}, PME~\cite{pme})
truncating the Coulombic real space contribution and the LJ part of the potential at $r_c=14$~\AA~.
For convenience we have assigned to both Na and Cl the mass of Ar. 
Whenever we consider a spherically truncated potential (ST LJ or NaCl), we add
the long range corrections to the LJ part of the potential to both energy and pressure. 
Thus our aim was to estimate the free energy of the untruncated potential rather than of the truncated system itself.

We simulate the systems running NVT Molecular Dynamics for about 10~ns with a
time step of 0.0025-0.005~ps (i.e four or two million time steps). For the LJ systems the time step
in reduced units is of about $\tau^*=0.001$, and we have discarted the configurations that
corresponds to the first 2~ns of each trajectory. 
Simulations were carried out in Intel(R) Xeon(R) CPU X5650 @ 2.67GHz processors,
which means about 255~ns/day for the 256 LJ atoms system, and 17~ns/day
for the 1372 LJ atoms system and the JC-NaCl system.
We stored configurations every 100 MD steps, which corresponds to approximately 10$^5$ independent configurations
per simulation run.
The temperature was keep constant using the velocity rescale
thermostat \cite{JCP_126_014101_2007} with a relaxation time of 1~ps.

\section{MC versus GROMACS/LAMMPS}

To confirm the validity of the free energy calculations obtained with the Molecular Dynamics
packages in an NVT ensemble, we simulate the same systems with a NVT Monte Carlo code. 
For the calculation of $\Delta$A$_2$ we need to compute the
mean square displacement of each particle from its lattice position.
Representing  the time evolution of the mean square displacement for the LJ/STS
computed via MC or GROMACS, we show that both
the average and its fluctuations are very similar with the two codes.

\begin{figure}[ht]
\centering
\includegraphics[clip,scale=0.3,angle=-0]{square_displacement.eps}
\caption{Mean square displacement of the LJ/STS system for
$\Lambda_E'$=27.08~$k_BT$/\AA$^2$ as obtained with MC and GROMACS from Einstein molecule
calculations.}
\label{fluctuaciones}
\end{figure}

It has been shown so far that the free energies obtained from MC and MD agree within their
respective error bar. This is true for all the results presented in Table I of the main paper.
However, we would like to point out that a somewhat larger deviation (of the order of 0.06 NkT) was observed
between MC and MD for a LJ ST system  truncated at $r_c=2.7\sigma$ (data not shown). 
The deviation is not large
but is clearly visible.  It is well known that some issues arise
when comparing the properties obtained from MC to those obtained from MD for a truncated potential\cite{demiguel06a,frenkel_smit}.
For instance, if one is interested in the pressure of the truncated potential itself then one should
add an impulsive correction  to the traditional virial expression to evaluated the pressure within 
a MC NVT run\cite{demiguel06a,frenkel_smit}. However  a long tail correction is added rather than 
an impulsive correction in MC runs since one is usually more interested in estimating the properties of the
untruncated potential rather than the truncated potential itself. There is no problem in including this
long tail correction also in MD runs, and in fact this was done in this work. However
the discontinuity of the potential at $r_c$ generates impulsive forces that can not be
handled in MD programs as GROMACS/LAMMPS that are based on a Taylor expansion of the particle position.
Thus, we recommend to use relatively large system sizes and large cut-off's (above $4.5\sigma$) for the
calculation of free energy of solids when using MD to minimize both system size effects 
and the problem of the discontinuity of the potential at the cutoff. 


\bibliographystyle{./apsrev}